# Exciton-Sensitized Second-Harmonic Generation in 2D Heterostructures

Wontaek Kim[1#], Gyouil Jeong[1#], Juseung Oh[1], Jihun Kim[1], Kenji Watanabe[2], Takashi Taniguchi[2] and Sunmin Ryu[1,3*]

[1]Department of Chemistry, Pohang University of Science and Technology (POSTECH), Pohang, Gyeongbuk 37673, Korea.

[2]Advanced Materials Laboratory, National Institute for Materials Science, 1-1 Namiki, Tsukuba, 305-0044, Japan.

[3]Institute for Convergence Research and Education in Advanced Technology (I-CREATE), Yonsei University, Seoul 03722, Korea

## ABSTRACT

The efficient optical second-harmonic generation (SHG) of two-dimensional (2D) crystals, coupled with their atomic thickness that circumvents the phase-match problem, has garnered considerable attention. While various 2D heterostructures have shown promising applications in photodetectors, switching electronics, and photovoltaics, the modulation of nonlinear optical properties in such hetero-systems remains unexplored. In this study, we investigate exciton-sensitized SHG in heterobilayers of transition metal dichalcogenides (TMDs), where photoexcitation of one donor layer enhances the SHG response of the other as an acceptor. We utilize polarization-resolved interferometry to detect the SHG intensity and phase of each individual layer, revealing the energetic match between the excitonic resonances of donors and the SHG enhancement of acceptors for four TMD combinations. Our results also uncover the dynamic nature of interlayer coupling, as evidenced by the dependence of sensitization on interlayer gap spacing and the average power of the fundamental beam. This work provides insights into how interlayer coupling of two different layers can modify nonlinear optical phenomena in 2D heterostructures.

**KEYWORDS**: second-harmonic generation (SHG), phase interferometry, two-dimensional heterocrystals, transition metal dichalcogenides, charge transfer, exciton sensitization

**Introduction**

The optical responses of two-dimensional (2D) semiconductors have recently attracted intensive research efforts for their distinctive characteristics and applications[1-4]. Because of the reduced dielectric screening and quantum confinement induced by the low dimensionality, they exhibit strong light-matter interactions[3, 4]. Unlike graphene with a non-zero bandgap, a monolayer of transition metal dichalcogenides (TMDs) exhibits substantial excitonic absorption and photoluminescence in the visible and NIR regions[3-5]. Besides linear optics, they also exhibit strong nonlinear optical (NLO) effects such as second[6, 7] and high-harmonic[8] generations, sum-frequency generation[9] and four-wave mixing[9]. Among those, the second-harmonic generation (SHG) that essentially doubles the input photon frequency is simplest and yet has allowed or revealed the determination of thickness and stacking[6, 10, 11], characterization and imaging of crystallographic orientation[6, 12-14], excitonic resonance[7, 15, 16], extraordinarily large susceptibilities[6, 17, 18], and interferometric determination of energy-dependent phase delays[19]. The SHG response of atomically thin crystals could also be modulated by physical[20], chemical[21] and electrical[22-24] means.

SHG has also seen intriguing phenomena and applications in homo and heterobilayers of natural and artificial origin. Its coherent nature leads to the even-odd alternation in the SHG intensity of 2H-type TMDs[6, 25] and vectorial superposition in randomly stacked multilayers[26, 27]. Heterobilayers produce elliptical SHG outputs because of finite inter-material phase differences associated with excitonic resonances complicating the superposition-based interpretation[19]. Using wavelength-dependent resonance enhancement, the stack angle could be determined from heterobilayers[28]. Also notably, substantial interlayer couplings have been detected in the SHG signals of bilayer systems[9, 29]. Moreover, the instantaneous response of the parametric process has allowed spectroscopic tracking of ultrafast charge transfer phenomena[30, 31], which complements conventional time-resolved spectroscopy based on dissipative processes like absorption[32] and emission[32]. The observed excitation-mediated interlayer transfers can be viewed as a photosensitization that bridges an optical excitation with subsequent physicochemical changes in a neighboring chemical entity via energy, charge or atom transfer[33]. During the sensitization process between a donor and an acceptor, the nonlinear electronic susceptibility responsible for SHG should undergo a change, which can be detected in a steady-state or time-resolved SHG method. Despite well-characterized excitonic resonances of 2D TMDs[5, 34], however, the effects of excitonic photosensitization on their NLO properties are not understood. Viewing the strong NLO signals and on-demand stackability of 2D TMDs, their heterobilayers are an ideal system to probe the solid-state photosensitization of SHG.

In this study, we reveal exciton-sensitized SHG in four types of TMD heterobilayers ($MoX_2$-$WY_2$, where X and Y can be either S or Se) using polarization-resolved spectroscopy and imaging as well as spectral phase interferometry. The results show significant changes in SHG intensity and phase differences for all combinations. The enhancement of SHG intensity in $MoX_2$ induced by the presence of an adjacent layer ($WY_2$) matched with excitonic absorption bands of the latter and vice versa, thus confirming mutual photosensitization of SHG in 2D heterobilayers. Furthermore, the dependence of the enhancement on the power of the fundamental beam suggests a dynamic nature of the interlayer coupling rather than a static one. Based on the effects of atom-thin spacers made of hexagonal BN (hBN), it was concluded that the sensitization is mediated by charge transfer rather than energy transfer. These findings demonstrate the excited-state dynamics affecting NLO properties of 2D heterocrystals and have implications for designing new photonic functions and engineering their performance.

**Results and Discussion**

***Energy-dependent modulation of SHG intensity.*** Four TMD heterobilayers including $MoSe_2$/$WS_2$ shown in Fig. 1a were fabricated using the conventional dry transfer method[19, 35, 36], allowing for both hetero-stack bilayer (2L) and monolayer (1L) areas. In the samples denoted by $AX_2$/$BY_2$, the latter ($BY_2$) contacts the substrate, while $AX_2$-$BY_2$ is a generic representation of heterobilayers regardless of stacking order. The crystallographic orientation of each monolayer was determined within 0.5 degrees by polarized SHG measurements as shown in Fig. 1b. The stack angle ($\theta_S$) between the armchair directions of the two layers was set to be 30 ± 1 degrees to make the resulting SHG polarizations perpendicular to each other[6] (Fig. 1c). Then, the SHG signals from one individual layer at the hetero-stack area could be exclusively obtained by blocking those from the other layer using a polarizer. Note that the detected signals correspond to the parallel component from the targeted layer. Figure 1d shows three raster-scan images of the SHG intensity ($I_{WS2}$) obtained from $WS_2$ by varying SHG photon energies ($2\hbar\omega$). To facilitate comparison, the intensity images of $MX_2$ were presented in enhancement factor ($EF_{MX2}$) by normalizing with respect to the average intensity from 1L $MX_2$. Interestingly, $I_{WS2}$ was strongly modulated by the presence of $MoSe_2$, and the degree of modulation was substantially varied with photon energy. At $2\hbar\omega$ = 3.10 eV (left), the $MoSe_2$ layer strongly suppressed $I_{WS2}$, whereas this trend was drastically reversed at $2\hbar\omega$ = 2.67 eV (right). At $2\hbar\omega$ = 2.85 eV, $MoSe_2$ had little influence on $I_{WS2}$. Similarly, $I_{MoSe2}$ was greatly affected by $WS_2$, and $EF_{MoSe2}$ ranged over 0.4 ~ 0.7 for the same photon energy set (Fig. S1a & b). The intensity modulation could be seen more quantitatively in the SHG spectra of $WS_2$ from 1L and 2L areas (Fig. 1e), where $EF_{WS2}$ varied from 40% to 200% with decreasing photon energy. The change in EF that can be attributed to the absorption of SHG signals by the $MoSe_2$ layer is minor because the absorptance of $MoSe_2$ is only 10 ~ 15% for $2\hbar\omega$ = 2.6 ~ 3.1 eV[5]. This suggests that interlayer couplings play a major role in the modulation of SHG intensity. As one previous SHG study suggested a negligible interlayer coupling in $MoS_2$-$WS_2$[19], we performed similar measurements for $MoS_2$-$WS_2$ as shown in Figs. S1c ~ e and found negligible intensity modulation for the same set of $2\hbar\omega$. Similar types of data obtained for two additional types of heterobilayers revealed a strong modulation for $MoSe_2$-$WSe_2$ (Figs. S1f & g) and a weak one for $MoS_2$-$WSe_2$ (Figs. S1h & i).

***Origin of SHG modulation.*** The results presented in Fig. 1 and Fig. S1 suggest that a specific type of interlayer interaction with significant energy dependence is responsible for the observed intensity modulation. To investigate this phenomenon further, we determined EF in a broader range of $2\hbar\omega$ (Figs. 2a ~ d) and compared it with the linear and nonlinear optical responses of individual monolayers. Figure 2a depicting $EF_{WS2}$ obtained for the $MoSe_2$/$WS_2$ heterobilayer shows that a net enhancement occurs for 2.40 ~ 2.75 eV with a maximum of 170% at 2.64 eV, which is consistent with Fig. 1. A similar enhancement was observed for the reversed stacking order, $WS_2$/$MoSe_2$ heterobilayer (Fig. S2). However, $EF_{WS2}$ of a different type of heterobilayer ($MoS_2$-$WS_2$) was almost constant near unity (Fig. 2b), suggesting that the variation of $EF_{WS2}$ is affected by both mating layers, $MoSe_2$ and $MoS_2$. Notably, $EF_{MoSe2}$ obtained for the $MoSe_2$-$WS_2$ sample (Fig. 2a) exhibited a distinctive modulation with its local maxima at energies (2.42 and 2.94 eV) different from $EF_{WS2}$. The change in $EF_{MoSe2}$ was similar when MoSe$_2$ was mated with WSe$_2$ (Fig. 2c). In contrast, $EF_{MoS2}$ was almost constant for high $2\hbar\omega$ (> 2.5 eV for Fig. 2b, > 2.8 eV for Fig. 2d) but increased significantly with decreasing $2\hbar\omega$ and reached ~10 for MoS$_2$-Wse$_2$ (Fig. 2d). Note that $EF_{MoS2}$ in Figs. 2b and 2d could not be measured for $2\hbar\omega$ < 2.4 eV because of the finite extinction ratio of the polarizer used to block signals originating from mating layers (see Supplementary Note 1 for detection limit). To extend the photon energy range while avoiding this complication, we fabricated heterotrilayers consisting of 1L MoS$_2$ and 2L WS$_2$, the latter of which gives two orders of magnitude smaller SHG signals than the former because of its centrosymmetry. As shown in Fig. S3, $EF_{MoS2}$ showed a clear maximum at 2.38 eV.

The SHG enhancement observed in heterobilayers is strongly linked to the excitonic resonances of the constituent layers. In MoSe$_2$-WS$_2$, for example, the maximum value of $EF_{WS2}$ occurred at the C resonance of the MoSe$_2$ layer, whereas $EF_{MoSe2}$ exhibited two local maxima at the C and B resonances of WS$_2$ (Figs. 2a and 2e). It is worth noting that the differential reflectance spectra obtained from 1L TMDs (Figs. 2e ~ 2h) were consistent with previous reports[5]. Similar cross-layer coincidences between SHG and absorption can be seen in MoSe$_2$-WSe$_2$ (Figs. 2c and 2g), where two local maxima in $EF_{MoSe2}$ can be associated with the C and D excitons of WSe$_2$[5, 37]. Interestingly, the two heterobilayers containing MoS$_2$ presented a similar trend but with some apparent anomalies. Most notably, the largest enhancement found for $EF_{MoS2}$ of MoS$_2$-WSe$_2$ (Fig. 2d) can be attributed to the C resonance of WSe$_2$ (Fig. 2h). However, high-energy exciton resonances (D of WSe$_2$ and C of MoS$_2$) did not lead to any noticeable SHG enhancement. In MoS$_2$-WS$_2$ (Figs. 2b and 2f), an enhancement was observed for low-energy excitons (B of WS$_2$), but not for high-energy resonances (C of both WS$_2$ and MoS$_2$), which will be explained below. It is also important to note that the enhancement requires intimate contact between the two constituent layers. In Fig. S4, we observed that the enhancement curves became more structured and reproducible when vacuum-annealed at 300 ºC for 2 hours. This is consistent with the fact that dry transfer leads to uncontrollable interfacial space filled with ambient hydrocarbons[38]. Although thermal annealing could partially drive out interfacial materials for improved interlayer contact[39], there was still a non-negligible degree of inhomogeneity, which caused spatial variations in EF values.

The correlation between the SHG enhancement and the optical absorption of mating layers is characteristic of photosensitization[33], more specifically, mediated by excitons.

However, the anomalous behaviors associated with high-energy excitons in the MoS$_2$-containing heterobilayers remain unanswered. To shed light on this issue, we determined the second-order susceptibility $\chi^{(2)}$ of the four individual TMDs (Fig. S5). To correct the spectral dependence of the employed optical setup, $\chi^{(2)}$ shown in Figs. 2i ~ 2l was normalized with respect to that of α-quartz as described in Methods. The four $\chi^{(2)}$ curves exhibit characteristic resonance behaviors[40], which could be related to the excitonic absorption peaks in Figs. 2e ~ 2h. However, we note that there is a non-negligible energy difference between them: for instance, the C resonance of MoSe$_2$ in Fig. 2e occurs at 2.60 eV, whereas its SHG maximum is centered at 2.74 eV (Fig. 2i). It can be attributed to distinctive selection rules for the two optical processes. It is known that dark excitons invisible in absorption can enhance SHG[15]. Alternatively, the difference may be due to the stack-induced energy shift of the excitonic states[41]. Despite the apparent discrepancy and lack of B resonances in $\chi^{(2)}$ of WS$_2$, careful comparison between absorption (Figs. 2e ~ 2h) and $\chi^{(2)}$ (Figs. 2i ~ 2l) leads to the observation that the exciton-sensitized SHG enhancement occurs only when the energy is far off from its own SHG resonance so that its own signals are sufficiently weak. For instance, EF$_{WS2}$ could be maximized near 2.64 eV by the C resonance of MoSe$_2$ (Fig. 2a) because I$_{WS2}$ is much weaker than I$_{MoSe2}$. EF$_{MoSe2}$ also showed two peaks where $\chi^{(2)}$ of MoSe$_2$ is small (Fig. 2i). This explanation is also valid for MoSe$_2$-WSe$_2$ (Fig. 2c). In contrast, the C resonance of MoS$_2$ failed to induce any SHG enhancement (Fig. 2b) near 2.87 eV, where I$_{WS2}$ is even stronger than I$_{MoS2}$ because of its own C and other resonances (Fig. 2j). The lack of SHG enhancement associated with high-energy excitons in MoS$_2$-WSe$_2$ can also be justified on the same ground.

In order to provide a systematic description of sensitized SHG, we consider heterobilayers consisting of a system-sensitizer pair (SYS-SEN), where the sensitizer transfers energy or charges to the system upon photoexcitation. The system then transforms into an excited or ionized state and subsequently returns to the ground state. During this process, $\chi^{(2)}_{SYS}$ governing I$_{SYS}$ may vary. However, we consider time-averaged changes because of the current steady-state measurements. When the areal density of transferred excitation is sufficiently low, the susceptibility of the system can be approximated as $\chi^{(2)}_{SYS} = \chi^{(2)}_{SYS}{}^o + \Delta\chi^{(2)}$, where $\chi^{(2)}_{SYS}{}^o$ and $\Delta\chi^{(2)}$ represent the unsensitized system and the sensitization-induced change, respectively[42]. We assume that $\chi^{(2)}_{SYS}{}^o$ is identical to $\chi^{(2)}$ of the system without the sensitizer. Then EF is given as $|\chi^{(2)}_{SYS}|^2/|\chi^{(2)}_{SYS}{}^o|^2$. Since EF is obtained by normalizing with respect to the SHG intensity of the system without mating layers, a small EF

value does not necessarily indicate insignificant interlayer sensitization if the denominator is sufficiently large. The nature of sensitization is explored below.

***SHG phase modulation.*** A comprehensive description of sensitized SHG requires an examination of complex susceptibility, wherein a non-zero phase value signifies an SHG phase delay with respect to the fundamental light[43]. Given the finite transfer time needed for sensitization and sensitization-induced alteration of the system's electronic structure, it is conceivable to anticipate modulation in the SHG phase as well as intensity. We used spectral phase interferometry to extract the phase information. As shown in Fig. 3a, a reference SHG beam ($\vec{E}_{REF}^{2\omega}$) coherently generated by α-quartz formed an interferogram ($\vec{E}_{SAM+REF}^{2\omega}$) with the SHG signals ($\vec{E}_{SAM}^{2\omega}$) from the sample. Systematic errors were minimized by taking the same optical path for $\vec{E}_{REF}^{2\omega}$ and $\vec{E}_{SAM}^{2\omega}$. The sensitization-induced modulation in the SHG phase of the system was defined as the phase difference ($\Delta\phi$) between SYS-SEN and SYS as illustrated in Fig. 3b. Detailed information on the spectral phase interferometry technique is available elsewhere[19].

Interferograms obtained at $2\hbar\omega$ = 2.48 eV for MoSe$_2$ with and without WS$_2$ as a sensitizer showed a significant $\Delta\phi$ (Fig. 3c), indicating that the SHG signal induced by photosensitization is significantly phase-delayed compared to the unperturbed one. However, a negligible $\Delta\phi$ was found for MoS$_2$ mated with a WS$_2$ sensitizer (Fig. 3d), and this contrast was consistently observed over the entire energy range (Figs. 3e and 3f). The same trend was also observed for $\Delta\phi$ of WS$_2$ as a system. It should be noted, however, that the energy dependence of $\Delta\phi$ for MoSe$_2$-WS$_2$ does not appear to correlate with the intensity enhancement shown in Fig. 2a. This discrepancy is partly because $\Delta\phi$ depends not only on the phase associated with $\Delta\chi^{(2)}$ but also on the relative contribution of $\Delta\chi^{(2)}$ to $\chi^{(2)}{}_{SYS}$. A more rigorous analysis is needed for a quantitative understanding of the observed $\Delta\phi$. We also note that the presence of the sensitizer layer in SYS-SEN may induce a finite phase change even without interlayer coupling. A multilayer-interference simulation based on optical constants as done by W. Hsu et al.[26] indicated that the phase change in the SHG signals of 1L TMD layers induced by the presence of another 1L TMD layer is 2 ~ 8 degrees for $2\hbar\omega$ = 2.2 ~ 3.2 eV[44]. $\Delta\phi$ in Fig. 3e is several times larger than the optically-induced phase variation and thus dominated by the sensitization-induced effects.

***Nature of SHG sensitization.*** As seen in 2L graphene[45] and MoS$_2$[3], intimate van der Waals (vdW) interactions can modify the electronic structures of 2D crystals. To investigate whether the observed SHG modulation originates from a static vdW perturbation, we performed a power dependence experiment for MoSe$_2$-WS$_2$. We selected two photon energies that generated maximum EF for each layer, namely 2.67 eV for EF$_{WS2}$ and 2.42 eV for EF$_{MoSe2}$ (Fig. 2a) with EF of the other layer being far from its maximum. Figure 4a shows that EF$_{WS2}$ increased from unity to 2.2 and reached a constant value of 1.7 with increasing the average power of the fundamental beam. However, EF$_{MoSe2}$ remained constant throughout all the power ranges. Neither WS$_2$ nor MoSe$_2$ exhibited any power dependence when separated from mating layers with 2L hBN. Therefore, we can conclude that the SHG modulation is not of a static origin. The enhancement behavior of MoSe$_2$ at 2.42 eV in Fig. 4b is also consistent with this conclusion: a noticeable power dependence was observed for EF$_{MoSe2}$, which showed a large enhancement (Fig. 2a), but not for EF$_{WS2}$.

Excluding the effect of static vdW coupling, various types of interlayer interactions may be responsible for exciton-sensitized modulation. These include charge[32, 46] or energy[47, 48] transfer from the sensitizer to the system, or excitonic electric field-induced transient renormalization of the electronic structure of the system[49]. As each interaction has a distinctive length scale for effective sensitization[32, 49, 50], we investigated how the modulation depends on the gap distance between the system and the sensitizer. To achieve precise control over the gap spacing, high-quality hBN crystals were used as a spacer (Fig. 5a and Fig. S6). Even-number layers of hBN (2, 6 and 8L) were selected to minimize the SHG contribution from the spacer itself[6]. As shown in Fig. 5b, the 2L spacer provided an average gap of 1.1 ± 0.33 nm, which is only slightly larger than the vdW thickness of 2L hBN[51]. Note that some degree of spatial inhomogeneities mostly arose from interfacial bubbles characteristic of the ambient dry transfer, as shown in the topographic AFM images of representative samples (Fig. S6). Surprisingly, the SHG enhancement of the 2L-spaced sample in Fig. 5c is monotonically flat near unity for both layers over the entire energy range. The modulation in phase is also very small (Fig. 5d). These responses differ from the unspaced case (Figs. 2a and 3e). Thicker spacers gave similar results (Figs. 5e, 5f, and S7), indicating that the photosensitization is extremely short-ranged and ineffective beyond 1 nm.

After examining the different types of interlayer interactions that could be responsible for the exciton-sensitized modulation, we have concluded that charge transfer (CT) depicted in Fig. 5g is consistent with all the observations. This conclusion is supported by several lines of evidence. First, the sub-nm short-range interaction is characteristic of tunneling-based transfer phenomena[52]. Indeed, CT-mediated reduction in SHG signals was noticeably mitigated by $Al_2O_3$ spacers of 1 nm thickness[9]. In contrast, Forster-type energy transfer (ET) between 2D donors and acceptors is effective even beyond several nm[47], and the effects of excitonic fields on the oscillator strength of $WS_2$ also reach more than a few nm[49]. Second, an ultrafast time scale is required for the sensitization because SHG is an instantaneous parametric process that does not involve a population of excited states. More specifically, the transfer time for the sensitization needs to be on the same order of magnitude or shorter than the temporal width of the fundamental pulse (~140 fs). We note that CT takes orders of magnitude shorter than ET: 50 fs for the former between $MoS_2$ and $WS_2$[32] vs. 4 ps for the latter between graphene and TMD[53]. Ultrafast CT has also been reported in several other 2D heterostructures, occurring within the temporal duration of fs laser pulses[32, 54, 55]. Recent time-resolved SHG[56] and 2D electronic spectroscopy[57] directly quantified the CT time of tens of femtoseconds. Ultrafast injection of photogenerated hot carriers has also been observed between plasmonic metals and 2D TMDs[58, 59]. This comparison supports that the sensitization is based on CT.

The current results shed light on the interlayer CT and subsequent sensitization mechanisms. In particular, the ultrafast photosensitization indicates that the CT process competes with excitons' radiative and nonradiative decay channels. Assuming that hot carriers relax to their band edges before CT, one can conclude that the bidirectional sensitization observed in the four type-II heterobilayers[60] involves transferring both types of charges because the band-edge electrons of the low-lying sensitizer are not energetically allowed to move to the other layer. Note that ultrafast transfer of both types of charges has been observed in TMD heterobilayers[32, 56]. Alternatively, photogenerated hot carriers can be directly transferred, as depicted in Fig. 5h. This differs from the conventional interlayer CT process briefly explained above. Because of excess energy, hot carrier transfer may occur in both directions regardless of its polarity and band alignment. It is worth noting that the sensitization efficiency reaches a plateau with increasing laser power[59]. This saturation behavior can be ascribed to exciton-exciton annihilation, which becomes the dominant relaxation channel at high exciton density[61, 62]. Alternatively, the transient potential barrier generated by transferred charge carriers could account for the saturation behavior. The resolution of this issue will require determining the transient charge density and gaining a quantitative understanding of how the density affects $\Delta\chi^{(2)}$.

**Conclusions**

In this study, we reported that exciton-mediated photosensitization can amplify SHG intensity by order of magnitude and modify SHG phase in TMD heterobilayers (MoX$_2$-WY$_2$; X = S or Se; Y = S or Se). To determine the origin of this phenomenon, we utilized polarization-resolved spectroscopy and imaging as well as spectral phase interferometry. For the four bilayer combinations of system-sensitizer, the intensity enhancement of the system coincided with the excitonic absorption bands of the sensitizer and vice versa. The dynamic nature of the mutual photosensitization via excitons was further revealed by the power dependence of the enhancement. We concluded that the sensitization is mediated by charge transfer rather than energy transfer, as indicated by the ultrashort-ranged sensitization. By elucidating the excited-state dynamics that affect the NLO properties of 2D heterocrystals, our findings may facilitate the design of new photonic functions and the optimization of their performance.

**Methods**

***Preparation of Samples.*** TMD (MoS$_2$, MoSe$_2$, WS$_2$, and WSe$_2$) and hBN samples were prepared by mechanical exfoliation of commercial (2D Semiconductors) and home-grown bulk crystals. Amorphous quartz was used as substrates after thorough cleaning using piranha solutions to avoid unwanted optical interference. Artificially stacked samples were prepared by transferring a top layer pre-exfoliated on a polydimethylsiloxane (PDMS) film on top of a bottom layer supported on a quartz substrate[19, 35]. For a targeted stack angle, the crystallographic orientation of each layer was first determined by polarized SHG measurements. Then, the dry transfer was made with the angular alignment of the top layer with respect to the bottom one[19]. The positional and angular errors for the targeted transfer were less than 2 μm and 1°, respectively. To improve the interfacial contact[19], heterobilayer samples were annealed in a vacuum at 300 °C for 2 hours.

***SHG Measurements.*** The measurements were performed with a home-built micro-SHG spectroscopy setup configured upon a commercial microscope (Nikon, Ti−U) as described elsewhere[19]. As a fundamental pulse, the plane-polarized output from a tunable Ti:sapphire laser (Coherent Inc., Chameleon) was focused on samples using a microscope objective (40×, numerical aperture = 0.60). The polarization of the fundamental beam was aligned parallel to the y-axis of the laboratory coordinates (Fig. 1c). The fwhm of the focal spot was 2.3 ± 0.2 μm. The pulse duration and repetition rate were 140 fs and 80 MHz, respectively. The backscattered SHG signals were collected with the same objective and guided to a spectrometer equipped with a thermoelectrically cooled CCD detector (Andor Inc., DU971P). To vary the direction of the fundamental's polarization with respect to exfoliated 2D crystals, they were rotated about a surface normal to their basal planes using a rotational mount with a precision of 0.2°.

For polarization-resolved measurements, an analyzing polarizer was placed in front of the spectrometer to select polarization components of interest. For the spectral phase interferometry, reference SHG pulses were generated in an in-line geometry (Fig. 3a) by focusing the fundamental beam at a z-cut α-quartz crystal of 100 μm thickness[19]. For photon-energy dependent measurements, the wavelength of the laser was varied in the range of 800 ~ 1060 nm. A Cassegrain-type reflective lens (Edmund Optics, 52×, numerical aperture = 0.65) was used to minimize the wavelength-dependent artifact. To prevent sample degradation, the average power density of the incident fundamental beam was maintained below 40 μW/μm$^2$ for the interferometric measurements and below 100 μW/μm$^2$ for the others.

***Determination of second-order susceptibility.*** The second-order susceptibility, $\chi_s^{(2)}$, of four types of 1L TMD samples was determined using a z-cut α-quartz crystal as a reference. According to previous reports[6, 40], $\chi_s^{(2)}$ is given as follows:

$$\frac{\left|\chi_s^{(2)}\right|/d_s}{\left|\chi_{q,11}^{(2)}\right|} = \frac{c}{4\omega[n(\omega) + n(2\omega)]}\sqrt{\frac{I_s}{I_q}}$$

, where $\chi_{q,11}^{(2)}$ is the second-order susceptibility of α-quartz along its 11 direction and has a constant value of 0.3 pm/V for the photon energy range of the measurements (2ℏω = 2.2 ~ 3.2 eV)[63]. $d_s$ is the thickness of 1L TMD samples[64], $c$ is the speed of light and $n$ is the refractive index of α-quartz. $I_s$ and $I_q$ are the SHG intensities of samples and quartz, respectively, normalized to the square of the fundamental beam's power. Our susceptibility values (Fig. S5) were consistent with those obtained from samples grown by chemical vapor deposition[40] within ~50%, and their overall energy dependencies were similar.

ASSOCIATED CONTENT

**Supporting Information.**

Additional data for energy-dependent SHG modulation; Dependence of SHG enhancement on stacking order; Sensitization of 1L $MoS_2$ by B exciton of 2L $WS_2$; Effects of thermal annealing on intensity EF; Determination of second-order susceptibility; Morphology of $WS_2$/nL-hBN/$MoSe_2$ heterostructures with hBN spacers of various thicknesses; Inhibition of SHG sensitization in $WS_2$/nL-hBN/$MoSe_2$ by hBN spacers of various thicknesses; Uncertainty of intensity EF induced by optical artifacts; Uncertainty of SHG enhancement factor (EF). This material is available free of charge via the Internet at http://pubs.acs.org.

This work was submitted to a pre-preprint server:

Wontaek Kim; Gyouil Jeong; Juseung Oh; Jihun Kim; Kenji Watanabe; Takashi Taniguchi; Sunmin Ryu. Exciton-Sensitized Second-Harmonic Generation in 2D Heterostructures. 2023, arXiv:2305.17512. arXiv. https://arxiv.org/abs/2305.17512 (May 27th, 2023)

AUTHOR INFORMATION

**Corresponding Author**

*E-mail: sunryu@postech.ac.kr

#These authors contributed equally.

**Author Contributions**

S.R. conceived the project. W.K., G.J. and S.R. designed the experiments. W.K., G.J., K.W. and T.T. prepared samples. W.K., G.J., J.O. and J.K. performed the spectroscopy experiments and analyzed the data. W.K., G.J. and S.R. wrote the manuscript with contributions from all authors.

NOTES

The authors declare no conflict of interest.

ACKNOWLEDGMENTS

This work was supported by the National Research Foundation of Korea (NRF-2020R1A2C2004865, NRF-2022R1A4A1033247) and Samsung Electronics Co., Ltd (IO201215-08191-01). G.J. acknowledges the support from the National Research Foundation of Korea (NRF-2021R1I1A1A01040734).

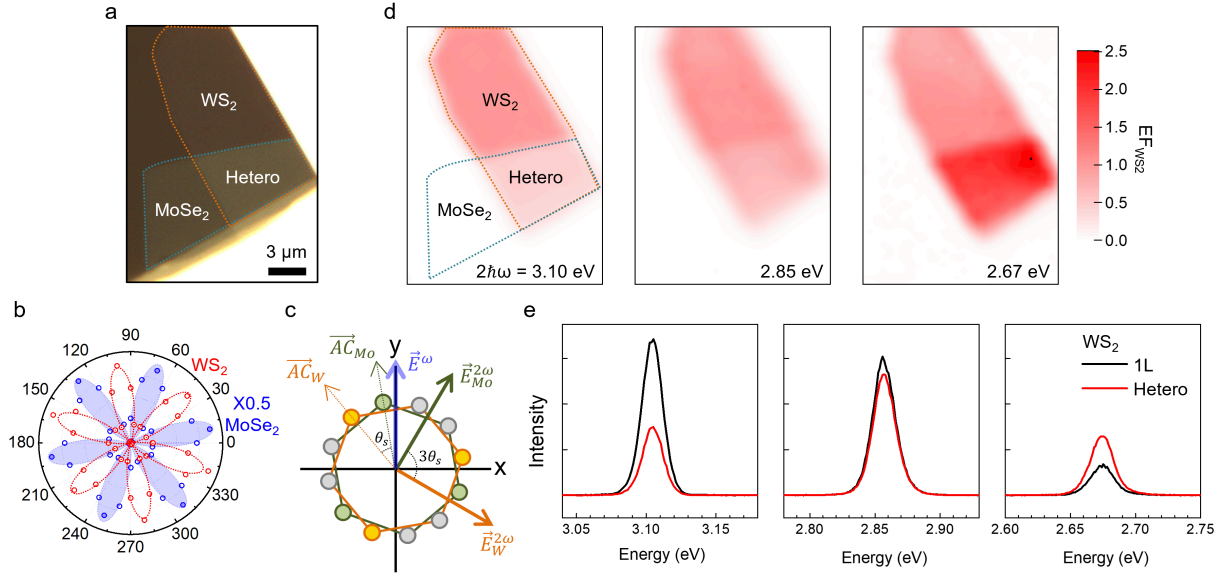

**Figure 1.** SHG intensity modulation in 2D MoSe$_2$/WS$_2$ induced by interlayer coupling. (a) Optical micrograph of artificially stacked MoSe$_2$/WS$_2$ heterobilayers. (b) Intensity polar graphs of 1L areas of (a), showing that the stack angle ($\theta_s$) is 30 ± 1°. (c) Scheme of polarization-selected SHG measurements. For a sample of $\theta_s$, defined with the armchair directions ($\overrightarrow{AC}_{Mo}$ and $\overrightarrow{AC}_W$) of individual layers, the SH fields from each layer ($\vec{E}^{2\omega}_{Mo}$ and $\vec{E}^{2\omega}_W$) generated by the fundamental field ($\vec{E}^\omega$) form $3\theta_s$ between their polarization directions. The SHG signals from each layer can be exclusively collected by blocking those from the other using a polarizer. To obtain $\vec{E}^{2\omega}_W$, for example, the heterobilayer samples were rotated so that $\vec{E}^{2\omega}_{Mo}$ was aligned along the x-axis and blocked. Also, note that the polarizer transmitted only the component of $\vec{E}^{2\omega}_W$ that is parallel to $\vec{E}^\omega$. (d) SHG enhancement images for WS$_2$ obtained with selected SHG energies of $2\hbar\omega$ = 3.10, 2.85, and 2.67 eV. The enhancement factor (EF) was obtained by normalizing with respect to the average intensity of 1L areas. (e) WS$_2$-selected SHG spectra of 1L WS$_2$ and hetero-stacked areas for the three energies.

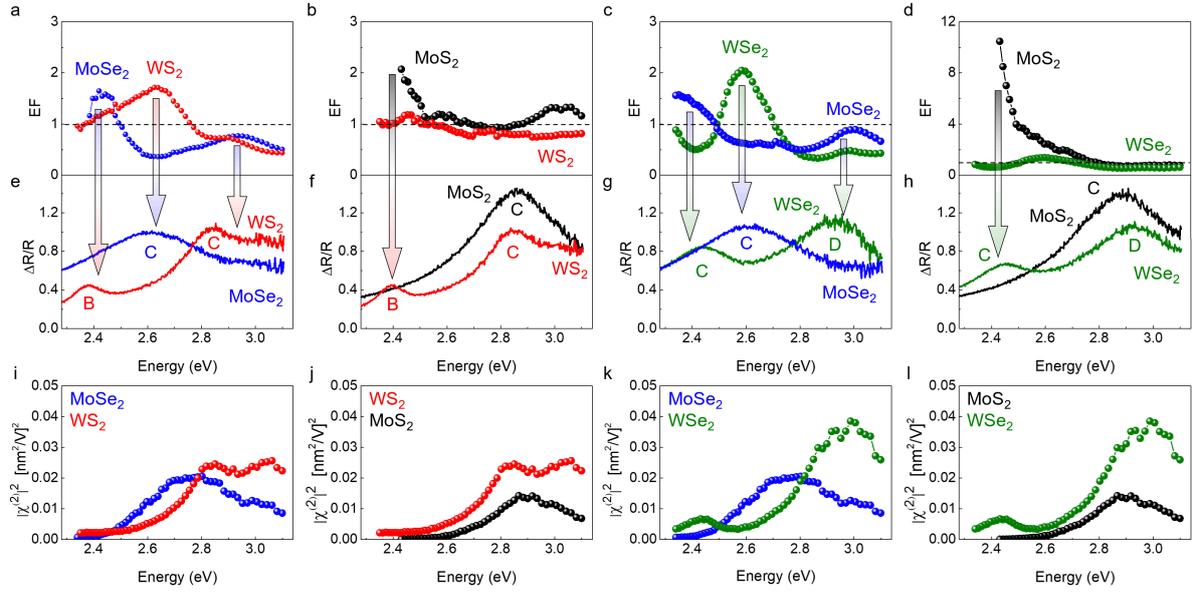

**Figure 2.** Energy-resolved SHG response of four types of heterobilayers ($MoSe_2$-$WS_2$, $MoS_2$-$WS_2$, $MoSe_2$-$WSe_2$, and $MoS_2$-$WSe_2$). (a ~ d) Intensity EF of the individual layers determined for four heterobilayers. (e ~ h) Differential reflectance ($\Delta R/R$) of four 1L TMDs shown in a pair corresponding to each heterobilayer in (a ~ d). Excitonic resonances are labeled with B, C and D. The vertical arrows denote the energetic correspondence between intensity enhancement and excitonic absorption. (i ~ l) Square of $\chi^{(2)}$ measured for four 1L TMDs shown in a pair corresponding to each heterobilayer in (a ~ d). The spectral sensitivity of the setup was corrected with respect to α-quartz crystals (Fig. S5).

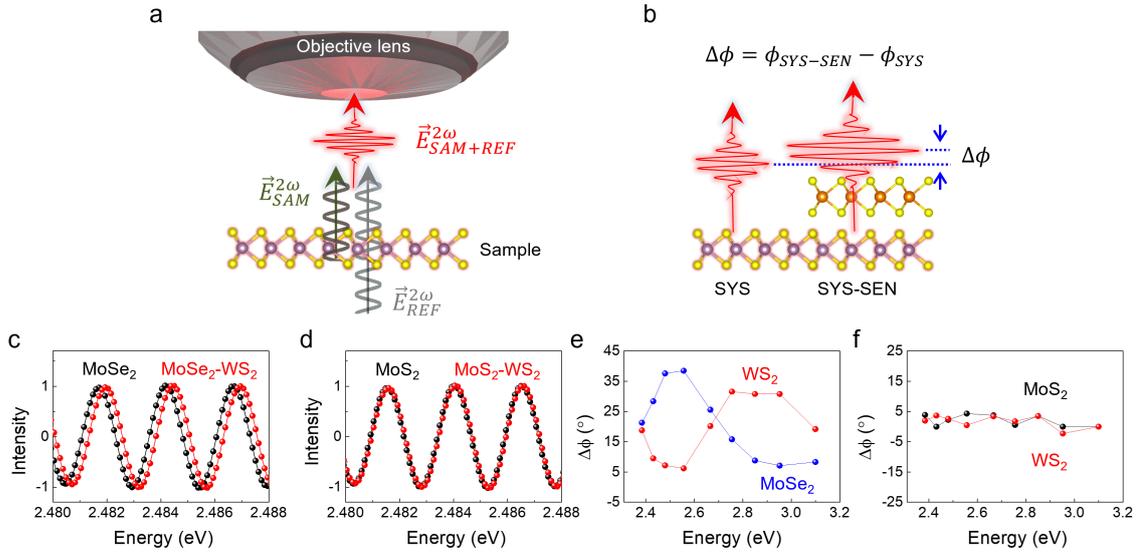

**Figure 3.** SHG phase modulation in heterobilayers. (a) Scheme of in-line spectral phase interferometry, where $\vec{E}^{2\omega}_{SAM}$ undergoes an interference with $\vec{E}^{2\omega}_{REF}$ generated by an α-quartz crystal. (b) Schematic representation of the sensitization-induced phase difference (Δ$\phi$) in a SYS-SEN heterobilayer. (c & d) Interferograms obtained selectively from $MoSe_2$ (c) and $MoS_2$ (d) layers from $MoSe_2$-$WS_2$ and $MoS_2$-$WS_2$ samples, respectively ($2\hbar\omega$ = 2.48 eV). Black and red symbols represent the data obtained from SYS and SYS-SEN areas, respectively. Slowly varying envelope functions were removed by Fourier transform to clearly show the oscillations (see Ref. 19). (e & f) Δ$\phi$ obtained for each layer of $MoSe_2$-$WS_2$ (e) and $MoS_2$-$WS_2$ (f) heterobilayer.

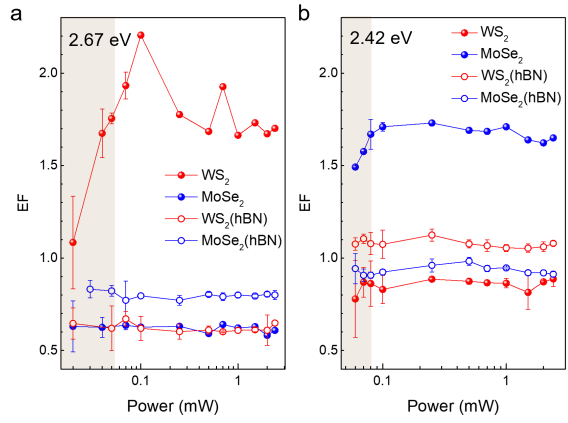

**Figure 4.** Power dependence of SHG intensity modulation. (a & b) EF for MoSe$_2$ (blue) and WS$_2$ (red) of MoSe$_2$/WS$_2$ (filled symbols) and WS$_2$/2L-hBN/MoSe$_2$ (empty symbols) obtained at $2\hbar\omega$ = 2.67 (a) and 2.42 eV (b).

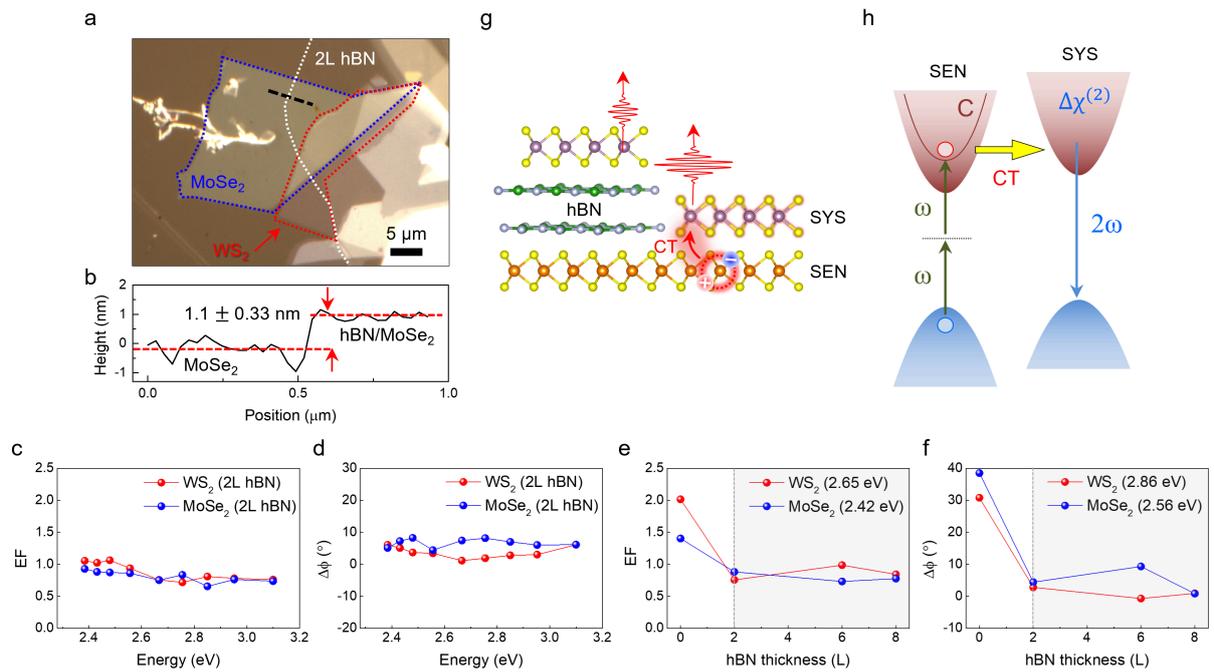

**Figure 5.** Exciton-sensitized SHG modulation and its gap dependence. (a) Optical micrograph of WS$_2$/2L-hBN/MoSe$_2$ heterostructure. Each layer is delineated with colored dotted lines. (b) AFM height profile for the spacer thickness obtained along the black dashed line in (a). (c & d) EF (c) and $\Delta\phi$ (d) obtained as a function of photon energy from the sample in (a). (e & f) EF (e) and $\Delta\phi$ (f) for MoSe$_2$ and WS$_2$ of WS$_2$/nL-hBN/MoSe$_2$ heterostructures (n = 0, 2, 6 and 8). Photon energies were selected for significant intensity enhancement and $\Delta\phi$ for each TMD of non-spaced samples. (g & h) Schemes of exciton-sensitized SHG modulation in real space (g) and energy-level diagram (h).

**TOC Figure**

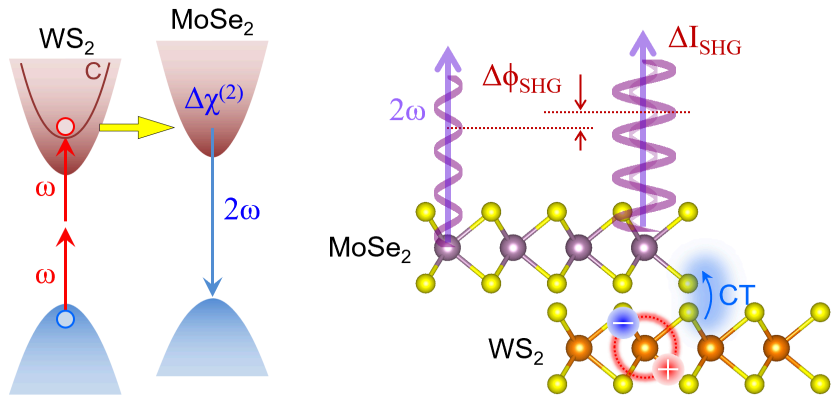